# THE INFLUENCE OF OROGRAPHY AND THE DIRECTION OF PREVAILING WINDS ON PRECIPITATION DISTRIBUTIONS

**Kochin A.V.**

Central Aerological Observatory

3 Pervomayskaya str., Dolgoprudny, Moscow region, 141707 Russia

Email: amarl@mail.ru

**Abstract**

The general circulation of the atmosphere (GCA) carries out a constant and unidirectional transfer of air masses, therefore its influence is manifested in the distribution of precipitation around the globe due to the occurrence of rain shadow behind mountain barriers. Over the territory of Russia, GCA manifests itself in the form of westerly winds, which cause a decrease in precipitation on the leeward side of the Ural Mountains and the Central Siberian Plateau. The contribution of the orographic component to the spatial variability of precipitation on average reaches 50-60% of the monthly precipitation amounts. Forecasting the magnitude of the orographic effect is close to predicting the transfer rates in GCA, the measurement of which has not yet been satisfactorily provided. A possible promising way to solve the problem is to develop special algorithms similar to those used for detecting Madden-Julian oscillations.

1. **Introduction**

The amount of precipitation depends on many reasons, including distance from the ocean, proximity of warm and cold currents, and terrain features. Most often, the formation of clouds with precipitation is associated with mesoscale processes such as cyclones. Average precipitation amounts are determined by the amount of water vapor in the atmosphere that can condense and form clouds. Condensation of water vapor followed by precipitation reduces the moisture content of the air mass. The resumption of precipitation is possible either after compensation of precipitation due to evaporation of water vapor from the surface, or by reducing the temperature of the air mass.

Turbulent transport is the main mechanism for the transfer of water vapor from the Earth's surface to the atmosphere. Thus, turbulent transport will reduce the humidity of the air mass with low evaporation from the surface in arid areas. Displacement of the air mass is the most likely way to resume precipitation. With the displacement of the air mass, the influence of mountains on the intensity of precipitation occurs, which is called the rain shadow, as a result of which the prevailing winds determine the position of arid areas relative to mountain ranges.

2. **The interaction of GCA and orography determines the amount of precipitation around the globe**

The regularities of the climatic distribution of precipitation are significantly determined by the orographic deformation of the prevailing wind. The orographic component makes the main contribution to the spatial variability of precipitation, and its total contribution on average reaches 50-60% of the monthly amounts. The general circulation of the atmosphere (GCA) carries out a constant and unidirectional transfer of air masses, therefore its effect is manifested in a decrease in the moisture content of the air from the leeward side of the mountain barrier. Thus, the interaction of the general circulation of the atmosphere and orography determines the average amount of precipitation around the globe [1, 2, 3, 4]. According to the prevailing ideas, the structure of air flows in the general circulation of the atmosphere looks in accordance with Fig.1.

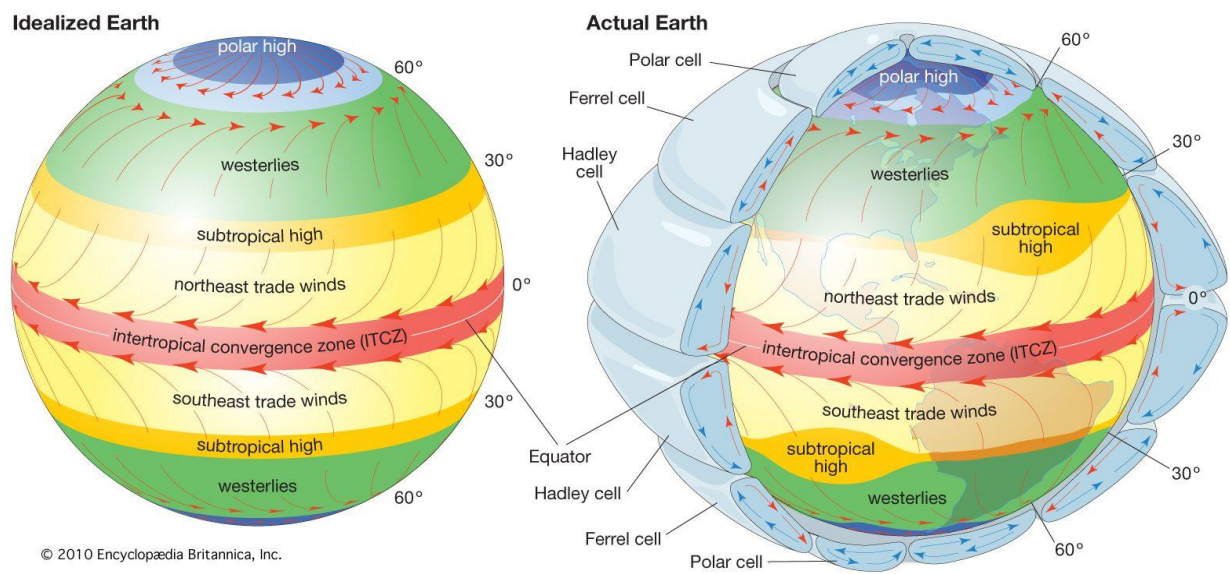

Fig. 1. The structure of air flows in the general circulation of the atmosphere. The idealized structure is shown on the left, the real structure is shown on the right.

Trade winds are located in the area from the equator to 30° latitude in both hemispheres. They deviate in a westerly direction. The area of latitudes from 35° to 65° is filled with westerly winds, which deviate in an easterly direction. A map of the distribution of annual precipitation is shown in Fig.2. The influence of the meridional component of the trade winds in the area from the equator to 30 ° latitude in both hemispheres is clearly manifested both in the formation of precipitation zones with high intensity and zones with reduced precipitation intensity.

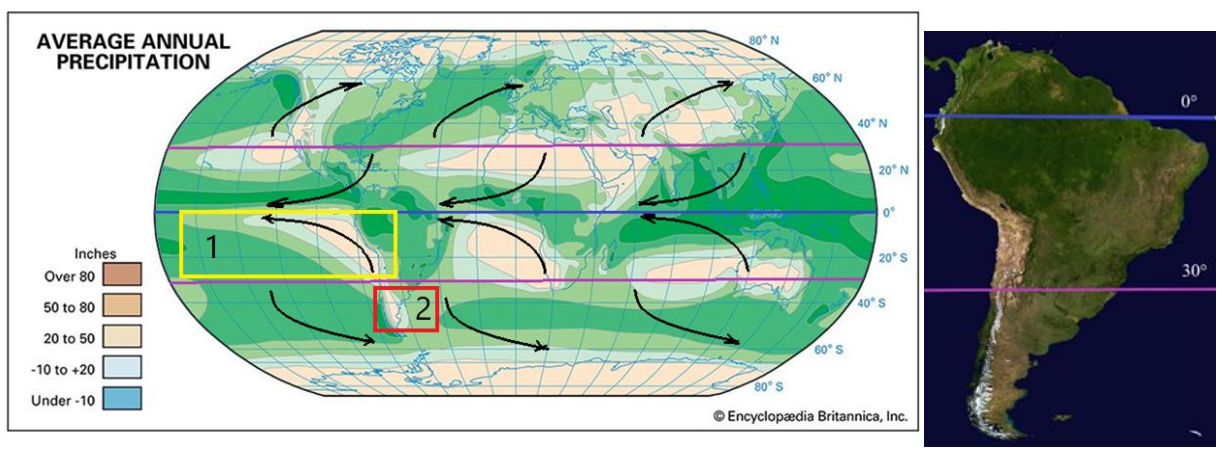

Fig. 2. Annual precipitation distribution map (left). The arrows show the directions of the air flow in the OCA according to Fig.1. The purple lines correspond to latitudes of 30°. The yellow and red rectangles highlight the areas of trade winds and westerly winds, respectively. A satellite topography map of South America is shown on the right.

The zonal distribution of annual precipitation has a noticeable heterogeneity, as shown in Fig.2. As an example, pronounced features in the region of South America are selected. In the yellow rectangle area 1, the minimum precipitation zone coincides with the direction of air flows in the trade winds (Fig.1), and in the temperate latitudes (highlighted in red rectangle in Fig.2) the zone of minimum precipitation also coincides with the direction of air flows in the westerly winds.

### 3. Precipitation distribution over the territory of Russia

Over the territory of Russia, GCA manifests itself in the form of prevailing westerly winds, blowing approximately between 35 and 65 degrees north latitude. Similarly, the influence of the mountain barrier is manifested in the distribution of annual precipitation over Russia (Fig. 3), which causes a decrease in precipitation from the leeward side of the Ural Mountains and the Central Siberian Plateau [3, 5] due to the influence of westerly winds.

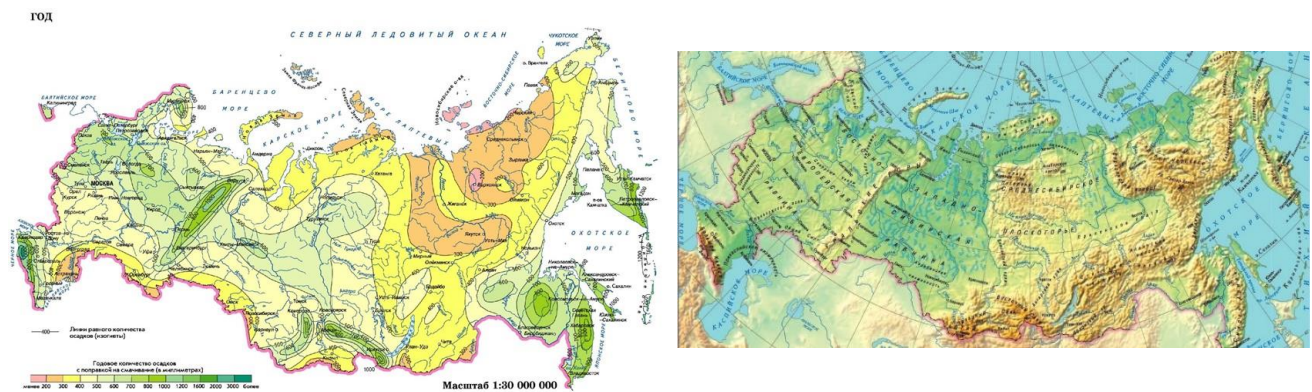

Fig. 3. Distribution of annual precipitation over the territory of Russia (left) and a relief map (right).

The distribution of clouds, snow cover, and wind velocity also have similar features (Fig.4). Everywhere the general picture remains, the influence of the Ural Mountains is especially pronounced.

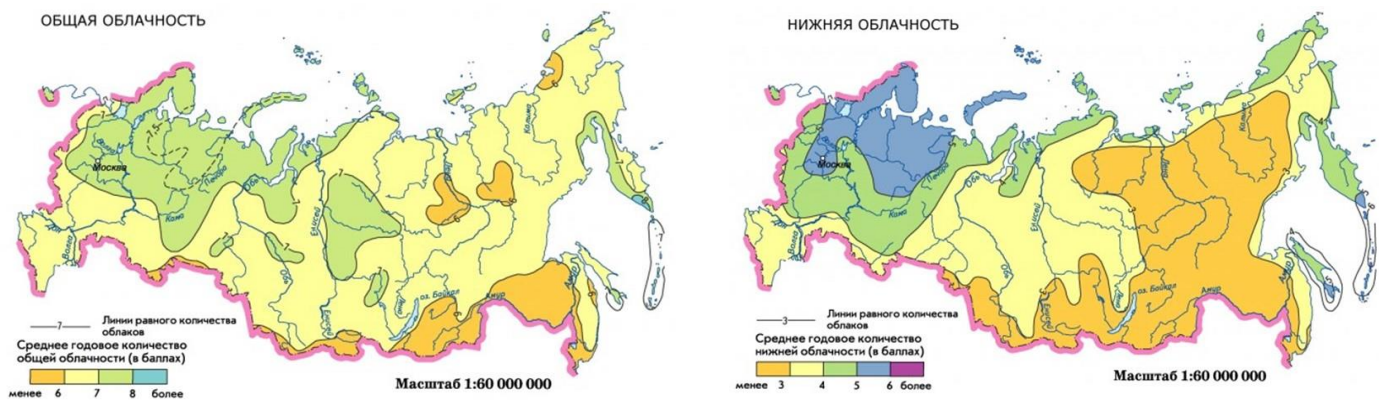

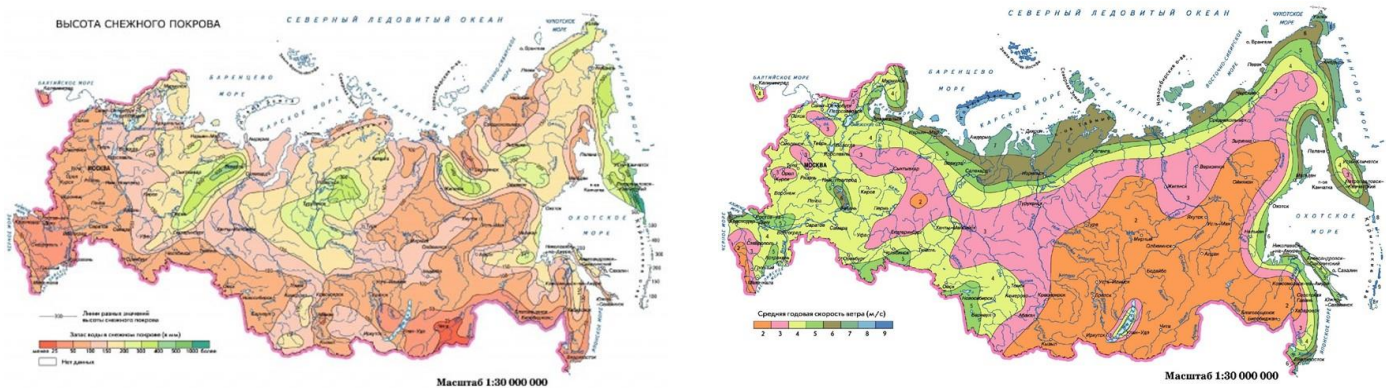

Fig. 4. Distribution of clouds, snow cover and distribution of wind speed. https://geographyofrussia.com/atmosfernye-osadki-v-rossii/.

The main contribution to spatial variability is made by the orographic component, and its total contribution on average reaches 50-60% of the monthly precipitation amounts. The influence of the orographic component must be taken into account for renewable energy sources, because it determines both the presence of clouds (solar power plants) and the intensity of winds (wind energy). The features of the processes, both by month and throughout the year, depend on the nature of the interaction of flows with a specific mountain barrier and the intensity of flows in the GCA, which is the subject of research in in Russia. [6]. The process of interaction of air flows in the atmosphere and a specific mountain barrier is the subject of research programs led by WMO [7, 8].

### 4. Convective cell behind the mountain barrier.

The distributions of precipitation and other atmospheric parameters, as follows from Figures 2, 3 and 4, are exposed to the mountain barrier for several thousand km. Such a distance of restoration of air parameters is difficult to explain even over land, not to mention the water surface. The explanation may be the occurrence of circulation in the form of a convective cell behind the mountain barrier. The air mass does not change during the passage of the mountain range, so the surface pressure remains unchanged. The density of warm air and the vertical pressure gradient after the barrier are lower, so at some height there is a zone of increased pressure (shown in green in Fig.5), which causes the appearance of a convective cell behind the mountain barter (Fig. 5). A temperature inversion is formed over the relatively cold surface of the land and, especially, the ocean, which blocks interaction with the surface and allows circulation to exist for a long time.

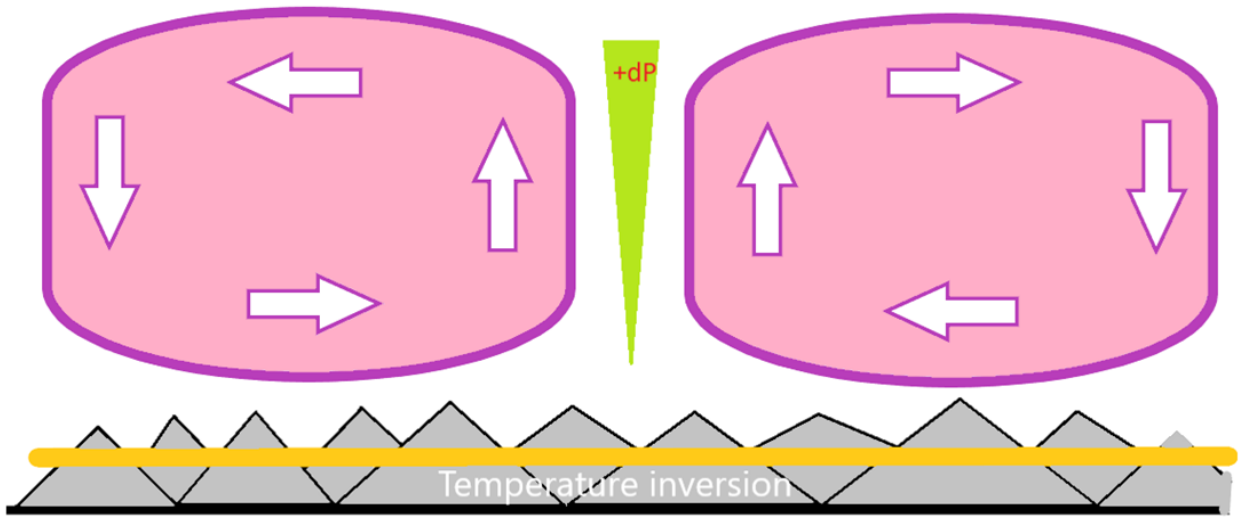

Fig. 5. Convective cell behind the mountain barrier. The circulation cell is shown in purple, the area of increased pressure is shown in green, and the inversion above the surface behind the mountain barrier is highlighted in orange.

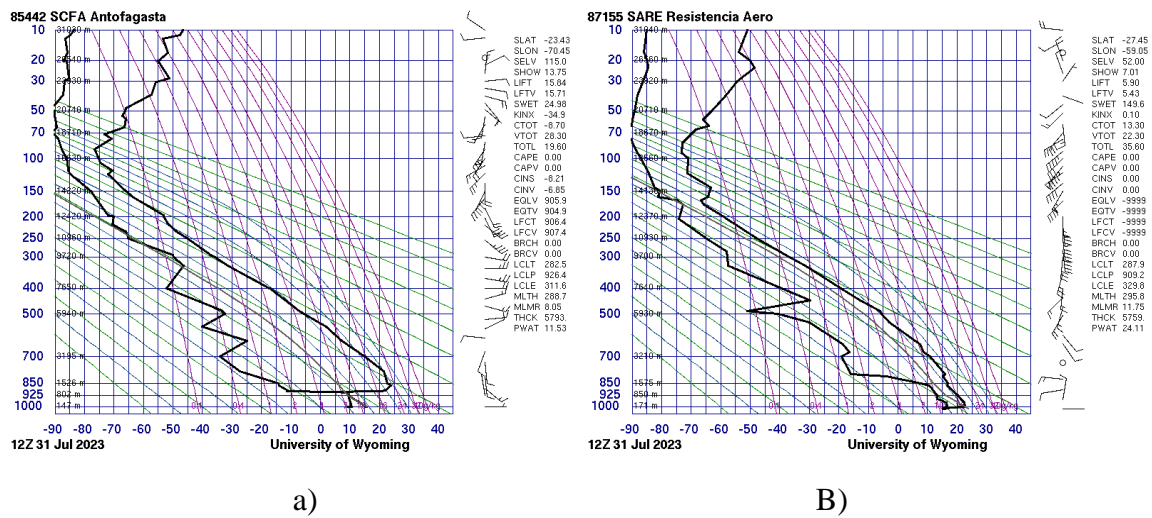

a)    B)

Fig.6. Upper-air sounding data on 07/31/2023 at 12.00 GMT on the territory of South America in the area of the mountain range. On the left (a) on the leeward side of the mountain barrier with inversion, on the right (b) on the windward side without inversion.

Three-hour amounts of precipitation and wind at the surface, as well as temperature and wind at 850 MB are shown in Fig.7. The date and time correspond to the date and time of the upper-air sounding.

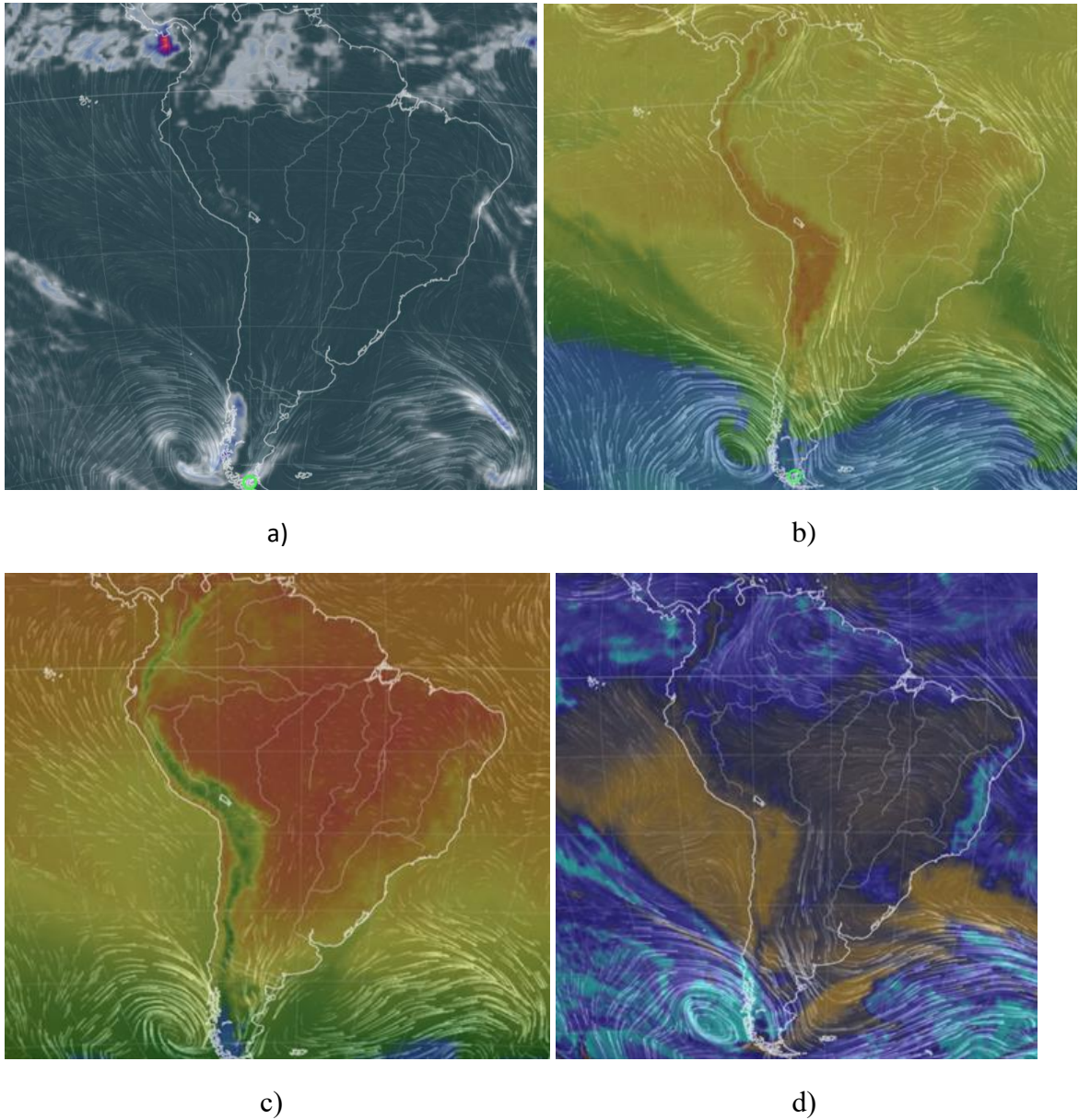

Fig.7. Three-hour amounts of precipitation and wind on the surface (a), temperature and wind at 850 hPa (b), temperature and wind on the surface (c) and relative humidity at 850 hPa on 07/31/2023 at 14.00 GMT (d) according to (https://earth.nullschool.net/ru /). Precipitation in Fig. 7a are highlighted in white.

At the level of 850 hPa (Fig.7 b), the area without precipitation (yellow) is 10 degrees warmer than the areas with precipitation (green). This effect is constantly observed in the area without precipitation. The surface temperature has no pronounced features (fig. 7c). Relative humidity at 850 HPa (Fig. 7 d) is noticeably lower in an area without precipitation.

**Measurement of the air flow velocity in the GCA**

The most important factor determining the average precipitation intensity for a time scale greater than the synoptic one is the intensity of flows in the GCA. However, it is precisely the measurement of the intensity of flows in the OCA that has not yet received enough attention. A friction layer (planetary boundary layer) with a thickness of about 1.5 km is located above the Earth's surface, in which the velocity of the ordered flow in the OCA decreases from units of m/s in the free atmosphere above the boundary layer to zero on the surface. The air flow velocity is measured at weather stations at a height of 10 meters above the surface. Consequently, the flow rate in the GCA cannot be determined on the basis of ground-based meteorological observations on the Earth's surface, because there it is practically zero. Upper-air sounding is the main source of information, although it does not provide the necessary accuracy for measuring wind direction at a flow rate of less than 2 m/s. Remote satellite methods estimate wind speed based on the displacement of inhomogeneities (clouds and similar inhomogeneities). These inhomogeneities arise from mesoscale processes and also move with them. Mesoscale processes have their own velocity of motion, which is an order of magnitude greater than the flow velocity in the GCA and has its own (generally asymmetric) distribution of the flow direction. Thus, the direct measurement of the air flow velocity in the OCA with existing devices and methods encounters noticeable difficulties.

5. **Forecasting the magnitude of the orographic effect**

The effect of the rain shadow is expressed in a decrease in precipitation from the leeward side of the mountain barrier. Predicting the magnitude of this effect is similar to predicting the transfer rate in GCA, which depends on the temperature difference between the equator and the poles (if the temperatures are equal, it is zero). A similar problem arises when studying a process close in meaning to the GCA, namely the Madden-Julian oscillation [9] (KMD), which can be traced in the fields of outgoing long-wave radiation, as well as the zonal component of the wind at the surface level of 850 and 200 hPa within the tropics, which is used to predict it. Information about the state of the KMD is part of the operational products of world meteorological centers (for example, the US National Meteorological Agency, the Tokyo Climate Center, the Bureau of Meteorology of Australia). Thus, to estimate the intensity of the GCA, one can use the already developed RMM index used in [9], or develop a new algorithm based on the difference in the zonal components of the wind below and above the tropopause [10]. A separate task, which has not yet been formulated, is the forecast of average precipitation amounts based on an assessment of the intensity of the GCA and the nature of the interaction of flows with a specific mountain barrier. For qualitative reasons, we can expect an increase in the ratio of precipitation intensity before and after the barrier (screening effect) with an increase in the flow rate in the GCA.

6. **Discussion**

The general circulation of the atmosphere carries out a constant and unidirectional transfer of air masses. For circulation to occur, it is necessary that regions with a multidirectional gradient exist in the atmosphere. Therefore, the pressure at the Earth's surface near the poles should be higher than the pressure at the equator. Only then do regions with a multidirectional pressure gradient form in the atmosphere. The air at the surface will shift towards the equator, and above a certain height with a zero gradient towards the pole. However, the observed surface pressure is almost constant, so due to the hydrostatic pressure gradient, the air can only shift towards the pole. The data on the distribution of precipitation shown in Figures 2 and 3 clearly indicate that the general circulation of the atmosphere exists and affects the distribution of precipitation. The creation of algorithms for determining the rate of transport in the general circulation of the atmosphere would be more successful if there was a theory explaining its existence.

However, the lack of a complete circulation theory is not a limitation for the development of algorithms for predicting the magnitude of the rain shadow. The accumulated volume of experimental data is able to ensure the development of effective algorithms. The success in predicting KMD is based precisely on the connection of the GCA with the fields of outgoing long-wave radiation and the zonal component of the wind at various levels.